# Two-dimensional electron gas formation in undoped In$_{0.75}$Ga$_{0.25}$As/In$_{0.75}$Al$_{0.25}$As quantum wells

F. Capotondi,[a), b)] G. Biasiol, I. Vobornik and L. Sorba.[a)]

*Laboratorio Nazionale TASC – INFM, 34012 Trieste, Italy.*

F. Giazotto.

*NEST –INFM and Scuola Normale Superiore, 56126 Pisa, Italy.*

A. Cavallini and B. Fraboni.

*INFM and Dipartimento di Fisica, Università di Bologna, 40127 Bologna, Italy.*

We report on the achievement of a two-dimensional electron gas in completely undoped In$_{0.75}$Al$_{0.25}$As/In$_{0.75}$Ga$_{0.25}$As metamorphic quantum wells. Using these structures we were able to reduce the carrier density, with respect to reported values in similar modulation-doped structures, to about 2–3×10$^{11}$ cm$^{-2}$ with mobilities of up to 2.15×10$^5$ cm$^2$ (V s)$^{-1}$. We found experimentally that the electronic charge in the quantum well is likely due to a deep-level donor state in the In$_{0.75}$Al$_{0.25}$As barrier band gap, whose energy lies within the In$_{0.75}$Ga$_{0.25}$As/In$_{0.75}$Al$_{0.25}$As conduction band discontinuity. This result is further confirmed through a Poisson-Schrödinger simulation of the two-dimensional electron gas structure.

[a)] Also at *Dipartimento di Fisica, Università di Modena e Reggio Emilia, 41100 Modena, Italy.*

[b)] Electronic mail: *capotondi@tasc.infm.it*.

# I. INTRODUCTION

Two dimensional electron gases (2DEGs) based on $In_xAl_{1-x}As/In_xGa_{1-x}As$ quantum wells (QWs) with high In concentration $x$ offer potential advantages over $Al_xGa_{1-x}As/GaAs$-based ones, both for fundamental physics studies and for device applications, due to some peculiar properties. First of all, the $In_xGa_{1-x}As$ effective electron mass $m^*$ is lower than the GaAs one, thus making $In_xGa_{1-x}As$ an attractive material for the realization of quantum devices. Since, in order to observe quantized phenomena it is necessary to obtain a sub band energy spacing $\Delta E$ of the order of $k_BT$, and $\Delta E$ is inversely proportional to $m^*$, a larger sub band spacing can be obtained by using a 2DEG channel layer with a lower $m^*$. Furthermore, by increasing the In content the Schottky barrier at the $In_xGa_{1-x}As$-metal contact becomes smaller, and is suppressed completely for $x \geq 0.75$, thus allowing the realization of highly-transmissive metal-semiconductor interfaces. This last peculiarity appears to be crucial in order to exploit the so-called proximity effect in superconductor/semiconductor hybrid systems,[1] or spin-related effects in ferromagnet/semiconductor devices.[2]

However, only heterostructures of one particular composition ($In_{0.52}Al_{0.48}As/In_{0.53}Ga_{0.47}As$) can be grown lattice-matched to a commercial binary substrate (InP). In order to maximize the flexibility in the choice of alloy composition, several authors[3-6] report the possibility of realizing metamorphic, almost unstrained $In_xGa_{1-x}As$ QWs by inserting "virtual substrates" consisting of InAlAs buffers with graded In composition grown on GaAs wafers. Using an InAlAs step graded buffer layer, a 2DEG with carrier density $n_s = 9.0 \times 10^{11}$ cm$^{-2}$ and electron mobility $\mu = 545\ 000$ cm$^2$ (V s)$^{-1}$ has been reported by Gozu et al.[4] in metamorphic modulation-doped $In_{0.75}Al_{0.25}As/In_{0.75}Ga_{0.25}As$ QWs grown on (001) GaAs substrate. Recently, Mendach et al.[6] reported a charge density down to $3.6 \times 10^{11}$ cm$^{-2}$ in modulation doped QWs with an inserted InAs channel with mobilities around 230 000 cm$^2$ (V s)$^{-1}$ in the dark.

A further reduction of the 2DEG carrier density, without a degradation of the mobility, would be desirable to study different classes of phenomena at experimentally accessible magnetic fields, when few Landau levels are filled. For example, Koch *et al.*[7] demonstrated a first-order phase transition from a spin-unpolarized paramagnetic state to a spin polarized ferromagnetic state in the Landau level $\nu = 2$ in a $In_{0.53}Ga_{0.47}As/InP$ QW ($n_s = 1.0\times10^{11}$ cm$^{-2}$).

In this article we show that the carrier density in metamorphic $In_{0.75}Al_{0.25}As/In_{0.75}Ga_{0.25}As$ QWs can be reduced ($< 3\times10^{11}$ cm$^{-2}$), with respect to previously reported values, by growing completely *undoped* epilayers, i.e., without any modulation-doping in the barriers. To our knowledge, this phenomenon has never been observed so far in this material system. We identified the charge transfer mechanisms to the QW as due to the presence of growth-related deep level states in the $In_{0.75}Al_{0.25}As$ barriers, with energies within the InAlAs/InGaAs conduction band discontinuity.

## II. EXPERIMENT

Samples were grown by solid source molecular beam epitaxy (MBE) on semi insulating (001) GaAs substrates. The 2DEG region (Fig.1) consisted of a 30 nm-wide InAlAs/InGaAs QW, grown at 470° C with a V/III beam flux ratio of 12 and placed from 50 to 250 nm from the surface in different samples. All samples were capped with a 10 nm protective InGaAs layer to prevent oxidation of the higher InAlAs barrier.

In order to reduce the threading dislocation density in the QW due to misfit dislocations generated at the heterointerface between mismatched layers, a 1.2 μm step-graded buffer layer of $In_xAl_{1-x}As$ ($x = 0.15–0.85$) was inserted below the 2DEG region to fit the GaAs lattice parameter to the $In_{0.75}Ga_{0.25}As$ one (Fig 1). The In concentration of up to 0.85 in the step graded buffer is chosen in such a way that the topmost part of the buffer has a lattice parameter corresponding to the unstrained $In_{0.75}Ga_{0.25}As$ due to the partial lattice relaxation of the buffer layers closer to the substrate.[8]

To avoid the formation of three-dimensional islands, the buffer layer was grown at a lower temperature (330 °C).[5,9] High-resolution x-ray diffraction was used to check the In concentration, and revealed also a strain relaxation of 95% in the 2DEG region.

As previously observed on InGaAs grown on GaAs substrates,[10] atomic force microscopy of our surface shows a well developed cross hatching pattern with a root mean square roughness of about 2-3 nm on a 25 μm × 25 μm scan field. The period of the roughness undulations in the [110] direction is in the sub-micrometric scale, smaller than in the [-110] direction.

The transport properties of the structures were evaluated in a standard 60-μm wide, 260-μm long Hall bar geometry defined with standard optical lithography and wet chemical etching. Ohmic contacts were obtained through e-beam metal evaporation and the thermal diffusion of a Ni-GeAu alloy into the sample. Hall bars were directed along the [-110] and [110] directions. Only in samples with mobility lower than 100 000 cm$^2$ (V s)$^{-1}$ did we observe an anisotropy of about 25% in the two directions, with higher mobility in the [-110] direction, likely due to the asymmetry in the surface roughness. [3,4]

Deep-level PICTS (Photo Induced Current Transient Spectroscopy) analysis of the In$_{0.75}$Al$_{0.25}$As barrier material were carried out with a Sula Tech.Inc. system, allowing temperature scans in the range 80K - 400K, with a heating rate of 0.2K/s. The emission rates varied from 5 to 2×10$^4$ s$^{-1}$ and applied biases ranged from 0.1V to 15V.

Photoemission spectroscopy was used to determine the Fermi level pinning at the surface used in the Poisson-Schrödinger simulation. Spectra were taken with a photon energy at 39 eV at the low-energy end station of the APE beamline at Elettra synchrotron of Trieste, equipped with a Scienta SES-2002 electron energy analyzer.[11]

III. RESULTS AND DISCUSSION

Figure 2 shows the transverse and longitudinal resistances at 1.4 K as a function of the perpendicular magnetic field up to 7 Tesla for a 2DEG placed at 130 nm from the surface with an electron mobility of $\mu =$ 215 000 cm$^2$ (V s)$^{-1}$. The longitudinal resistance shows the break of spin degeneracy at about 2.5 T, and its vanishing at magnetic fields that correspond to even integer quantum filling factors ν ≥ 4 demonstrates the formation of a 2DEG, without any additional parallel conducting channels in the structure. The Fourier analysis of low-field oscillations (< 2 T) shows a single frequency (see inset), thus revealing the presence of only one populated electronic sub-band in the QW, with a carrier density of $2.84\times10^{11}$ cm$^{-2}$, in agreement with the classical Hall measurement.

We observe that, using the same structure, the 2DEG transport characteristics are reproducible from sample to sample for a given set of growth parameters, within an experimental error of 10%. Furthermore, the carrier density (ranging from ~$1.5\times10^{11}$ cm$^{-2}$ to >$4\times10^{11}$ cm$^{-2}$) and mobility can be controlled by changing the growth conditions such as temperature and III/V ratio. Such investigations are still under progress and will be reported later.

In order to understand the origin of the electronic charge in the QW, we investigated the electrical properties of an undoped 500 nm-thick In$_{0.75}$Al$_{0.25}$As metamorphic layer grown in the same conditions as the QW structure reported above. No measurable free carrier density was detected by Hall measurements both at room temperature and at 1.4 K, thus ruling out the presence of unintentional shallow donor impurities inside the In$_{0.75}$Al$_{0.25}$As barrier as a possible source for the 2DEG formation.

The 2DEG could however be induced by deeper level donor states (not measurable by conventional transport techniques) provided that their energy positions lie within the In$_{0.75}$Al$_{0.25}$As/In$_{0.75}$Ga$_{0.25}$As conduction band discontinuity $\Delta E_c$ (estimated to be 0.24 eV from data in Ref. [12] and [13]). In this case, since the levels would lie higher in energy with respect to the In$_{0.75}$Ga$_{0.25}$As conduction band minimum, electrons could be transferred into the well, as in modulation-doped structures. To detect such deep levels we used the PICTS technique. Contrary to the more widely used Deep Level Transient Spectroscopy, PICTS allows one to identify deep electrically active levels, and to measure their activation energy and capture cross section, even in highly resistive materials.[14,15] Thus we characterized, for the first time to our knowledge, the deep level structure on *intrinsic* In$_{0.75}$Al$_{0.25}$As layers, without the intentional introduction of Si dopants, that could induce donor related levels, such as DX centres.[16]

A PICTS spectrum of a 500 nm-thick $In_{0.75}Al_{0.25}As$ layer, taken with $\lambda = 670$ nm and an applied bias of 5 V, is shown in Fig. 3. The peaks corresponding to deep levels inside the $In_{0.75}Al_{0.25}As$ band gap, were labelled A, B, C, D, E, F, G, H and their activation energy and apparent capture cross section are reported in Fig. 4. The error associated with the deep level apparent activation energy, calculated from a chi-squared fitting procedure to each data set of the Arrhenius plot in Fig. 4, is about 8%. The reported energies are not assigned to a specific band edge (conduction or valence) since PICTS detects without distinguishing electron and hole levels, i.e. levels that emit carriers to the conduction and to the valence band. Several defects have been reported in the literature for MBE-grown $In_{0.48}Al_{0.52}As$ layers,[17,18] with a wide spread of activation energies and defect concentrations depending on the growth conditions. In particular, our PITCS spectra are similar to those observed using DLTS and admittance spectroscopy by Luo *et al.*,[17] who found three wide peaks in the same temperature range as ours, that could thus represent a convolution of our sharper structure. In that work, the defects increased up to $1–2\times10^{16}$ cm$^{-3}$ by lowering the growth temperature from 565 °C to 500 °C and have been related to arsenic defects.

The quite large number of defects observed in our high resistivity InAlAs samples suggests the hypothesis of a high degree of compensation within the material. The role played by each level in the compensation process still needs to be clarified, even if the high resistivity suggests the presence of both deep acceptor and donor levels.

If we assume that the activation energies of levels A and B are both assigned to the conduction band, their energy position with respect to the $In_{0.75}Al_{0.25}As$ conduction band minimum (0.12 and 0.17 eV, respectively) is smaller than $\Delta E_c$. They are therefore good candidates for the formation of the 2DEG. To verify this hypothesis we simulated the dependence of the 2DEG density as a function of the InAlAs/InGaAs QW distance from the surface with a Poisson-Schrödinger solver.[19] We compared the result with the experimental data obtained on a set of samples grown in the conditions reported above. In the computation we used the energy gap and the effective electron mass reported by Vurgaftman *et al*[13] and the conduction band discontinuity $\Delta E_c$ from the data of Ref. [12] and [13].

We used photoemission spectroscopy to estimate the position of the Fermi level at the surface.[20] We measured independently the valence band edge on a 10 nm $In_{0.75}Ga_{0.25}As$ layer grown on an $In_xAl_{1-x}As$ buffer, and the Fermi edge on a polycrystalline Ag in electrical contact with the sample. The data were taken in the temperature range 100-300 K. From the spectra at T= 100 K we find that the Fermi level is pinned 30 meV below the conduction band minimum.

As shown in figure 5 (circles) the measured carrier density at 1.4 K, evaluated as the average value of a set on Hall bar devices, is roughly saturated at about $3.1 \times 10^{11}$ cm$^{-2}$ for QW depths greater than 150 nm. However, moving the QW towards the free surface, $n_s$ decreases to values lower than $2 \times 10^{11}$ cm$^{-2}$ at 50 nm, and the corresponding mobility decreases to about $7.7 \times 10^4$ cm$^2$ (V s)$^{-1}$. These effects are due to the formation of surface states that pin the Fermi level at the surface, thus depleting the shallower 2DEGs. This result is in contrast to Ref. [21] where no evident dependence of the density on the 2DEG depth was observed. This difference could be ascribed to the presence of a modulation-doping layer in those structure that could hinder surface depletion effects.

In figure 5 we plot the simulated 2DEG density in the cases in which only the A level or the B level is assumed to be a donor (long-dash line and dash-dot line respectively), and in the case in which both A and B are assumed to be donors (short-dash line). Since the PICTS technique can not measure the deep level densities $N_{dd}$, we have used in the simulation the values that best reproduce the experimental data in each of the three cases considered (A only: $N^A_{dd} = 2.3 \times 10^{16}$ cm$^{-3}$, B only: $N^B_{dd} = 4.4 \times 10^{16}$ cm$^{-3}$, A and B: $N^A_{dd} = 1.2 \times 10^{16}$ cm$^{-3}$ and $N^B_{dd} = 2.0 \times 10^{16}$ cm$^{-3}$). Figure 6 is an example of a simulated band profile for a QW located at 130 nm from the free surface, showing that the carrier density is completely contained in the QW with only one populated electronic sub-band, in agreement with magnetotransport measurements.

A good agreement with the experimental data can be obtained using only the A level, in particular for the 2DEGs close to the surface, even if it overestimates slightly the 2DEG density of the deepest 2DEGs. The B level can not reproduce the experimental data for any estimated density $N^B_{dd}$: $n_s$ saturates at a value lower than in the experimental data, and in the range 70–120 nm it overestimates the observed density, so we can exclude the B level as the only main source for 2DEG formation. The simulation using both levels, where the chosen ratio of ~1.7 between their densities is that of the PICTS peaks, (Fig. 3 inset), overestimates $n_s$ for the shallower 2DEGs, but gives a good agreement for the saturated value. This suggests that the B level could have some function in the 2DEG formation when surface depletion effects become negligible.

The possibility to fit the experimental data of a wide set of samples using a single concentration of deep donor levels as a fitting parameter demonstrates that their concentration does not change significantly among the different growths, and thus the reproducibility of our approach for the fabrication of low density 2DEGs in $In_{0.75}Al_{0.25}As/In_{0.75}Ga_{0.25}As$ QWs.

## IV. CONCLUSION

In conclusion, we showed that the carrier density in $In_{0.75}Al_{0.25}As/In_{0.75}Ga_{0.25}As$ 2DEGs can be reduced in a reproducible way to $2-3\times10^{11}$ cm$^{-2}$ by growing completely undoped heterostructures. Through PICTS measurements in combination with a Poisson-Schrödinger solver we established that 2DEG population takes place through a charge transfer from deep level donor states present in the $In_{0.75}Al_{0.25}As$ barrier with an activation energy that lies 0.12–0.17 eV below the conduction band minimum, within the $In_{0.75}Al_{0.25}As/In_{0.75}Ga_{0.25}As$ conduction band discontinuity.

Figure Captions

Figure 1. Schematic illustration of the In$_{0.75}$Al$_{0.25}$As/In$_{0.75}$Ga$_{0.25}$As QW structure grown on (001) GaAs substrate.

Figure 2. Longitudinal and transverse resistances of a 2DEG in an In$_{0.75}$Al$_{0.25}$As/In$_{0.75}$Ga$_{0.25}$As QW measured at T=1.4 K as a function of the perpendicular magnetic field *B*. Inset: Fourier analysis for *B* < 2 T.

Figure 3. PITCS spectrum from 80 K to 400 K of a metamorphic 500 nm-thick In$_{0.75}$Al$_{0.25}$As layer. Inset: gaussian fit of energy level A, B and C in the 90-140 K temperature range.

Figure 4. Arrhenius plot of the activation energies for each deep level. Best-fit values of the activation energies and capture cross sections are reported for each level.

Figure 5. Dependence of the carrier density $n_s$ inside the In$_{0.75}$Al$_{0.25}$As/In$_{0.75}$Ga$_{0.25}$As QW as a function of the QW depth. Cirles: experimental data. Long-dash line: simulation by taking into account only the A level. Dash-dot line: the same with only the B level. Short-dash line: the same with both the A and the B levels.

Figure 6. Simulated bands and density profiles for a 130 nm-deep In$_{0.75}$Al$_{0.25}$As/In$_{0.75}$Ga$_{0.25}$As QW using only deep level A with a density of $2.3\times10^{16}$ cm$^{-3}$.

| | | |
|---|---|---|
| | In$_{0.75}$Ga$_{0.25}$As | 10 nm |
| Barrier | In$_{0.75}$Al$_{0.25}$As | 40 - 240 nm |
| Well | In$_{0.75}$Ga$_{0.25}$As | 30 nm |
| Barrier | In$_{0.75}$Al$_{0.25}$As | 100 nm |
| In$_x$Al$_{1-x}$As  Buffer Layer $x= 0.15 \div 0.85$ | | 1.2 μm |
| GaAs substrate | | |

Figure 1.

Capotondi et al., J.V.S.T. B

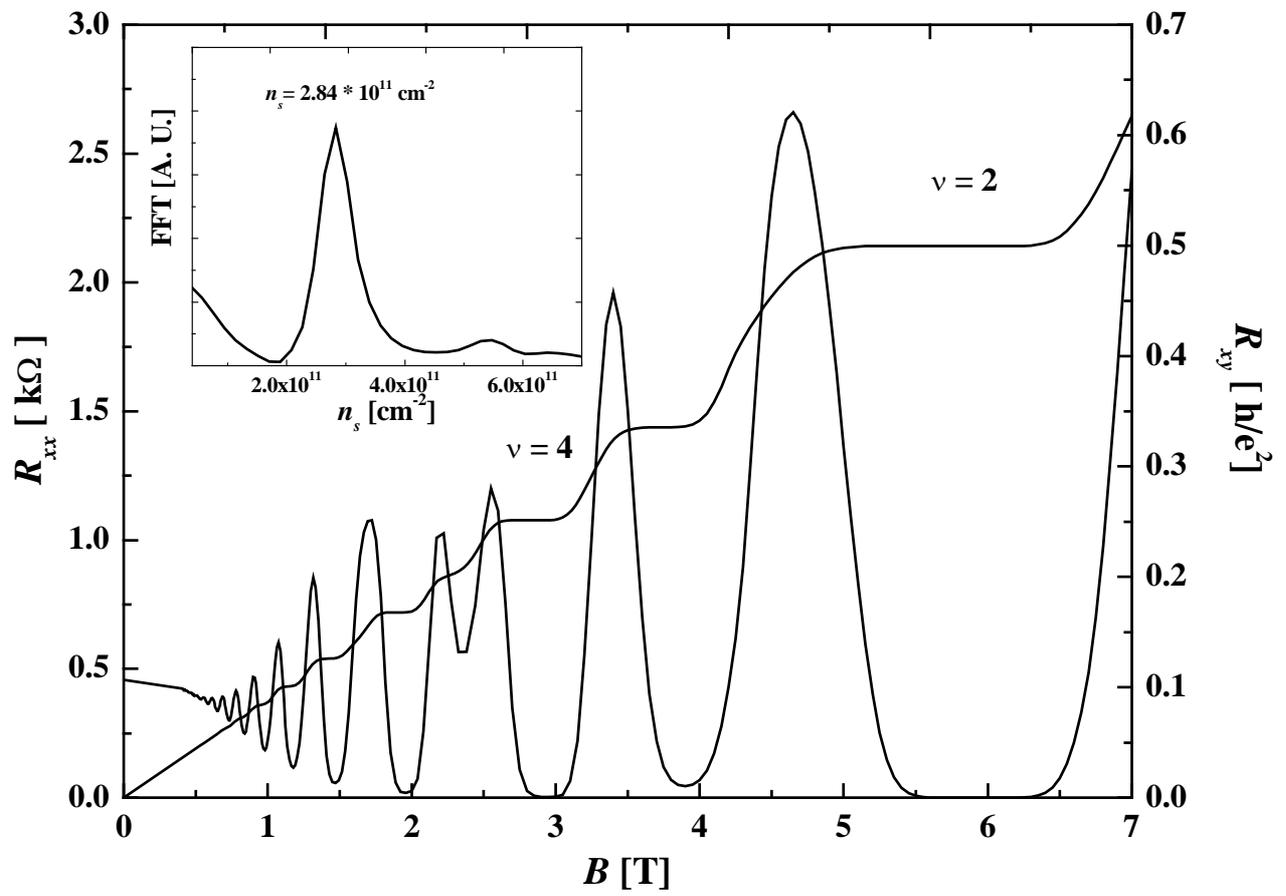

Figure 2.

Capotondi et al., J.V.S.T. B

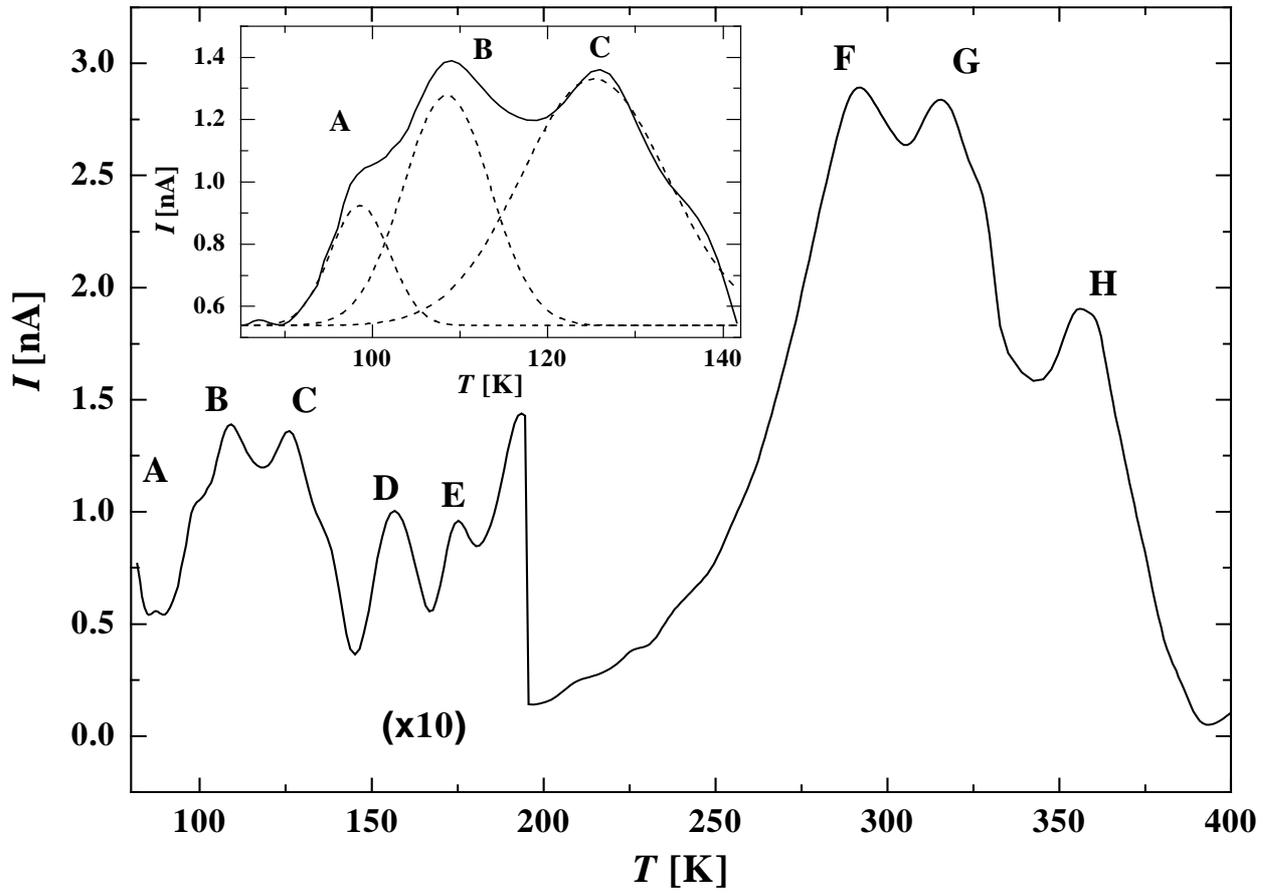

Figure 3.

Capotondi et al., J.V.S.T. B

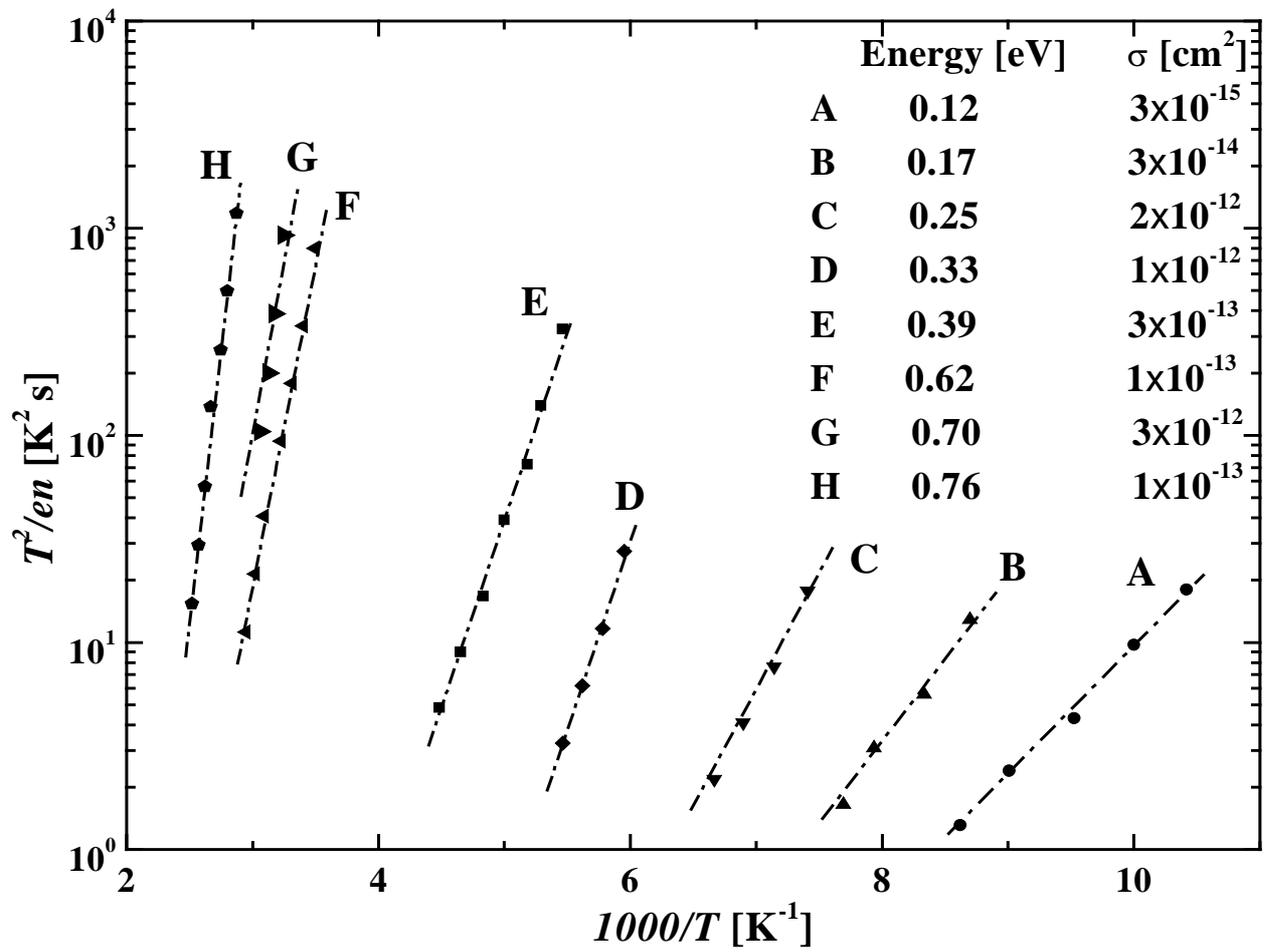

Figure 4.

Capotondi et al., J.V.S.T. B

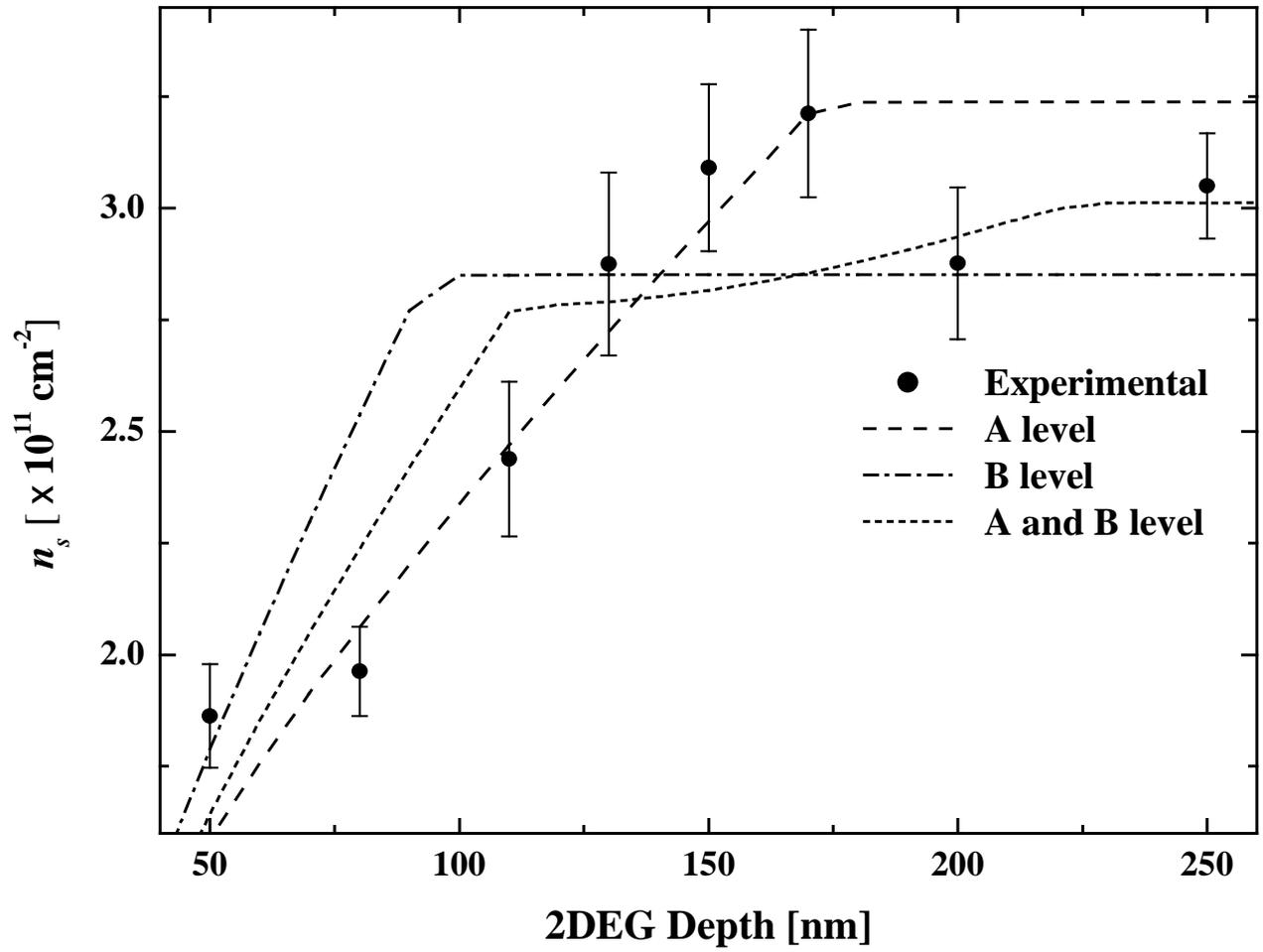

Figure 5.

Capotondi et al., J.V.S.T. B

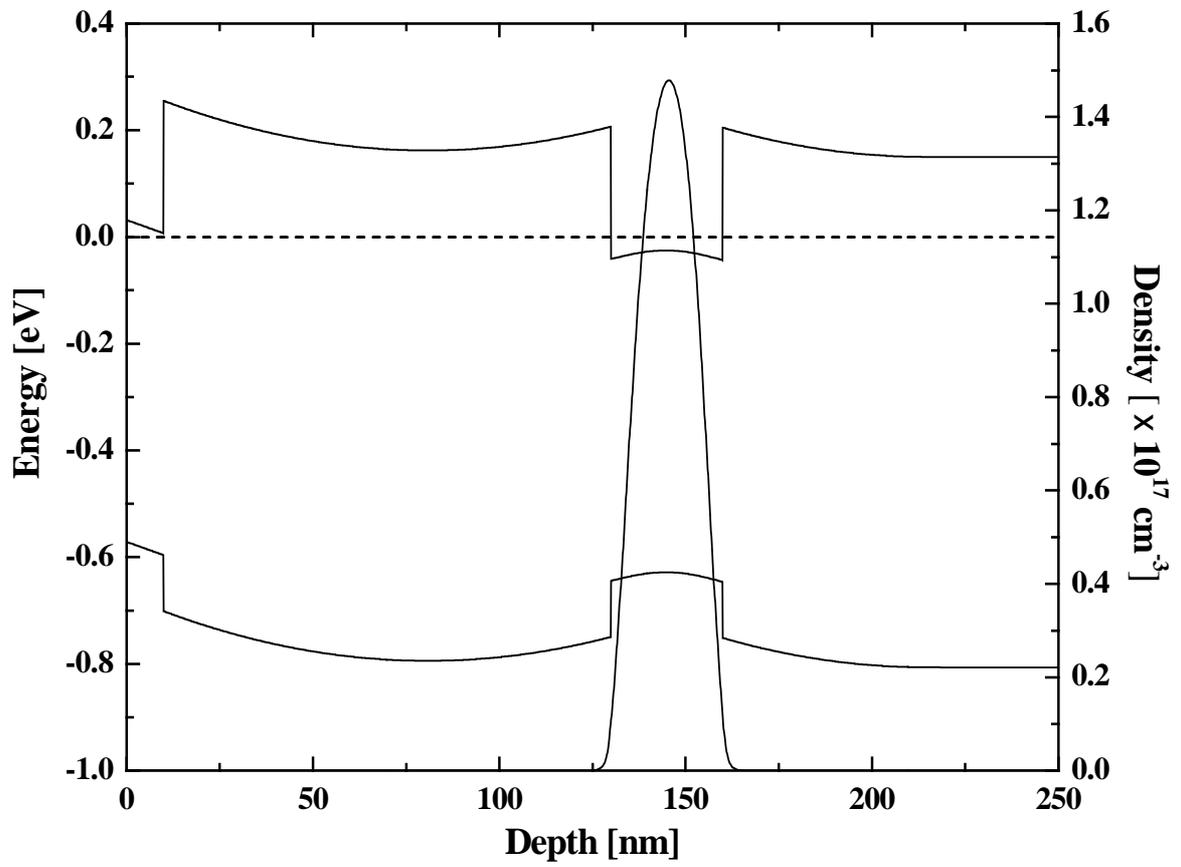

Figure 6.

Capotondi et al., J.V.S.T. B